\documentclass[aps,prx,10pt,twocolumn,superscriptaddress,amsfonts,amssymb,amsmath,floatfix]{revtex4-2}
\usepackage{microtype}
\usepackage{graphicx}
\usepackage{bm}
\usepackage{appendix}
\usepackage{hyperref}
\hypersetup{
    colorlinks=true
}
\usepackage[capitalise]{cleveref}
\usepackage{xcolor}
\usepackage{soul}

\let\oldhat\hat
\renewcommand{\vec}[1]{\mathbf{#1}}
\renewcommand{\hat}[1]{\mathbf{\oldhat{#1}}}
\DeclareRobustCommand{\l}{\left}
\DeclareRobustCommand{\r}{\right}

\begin{abstract}
The Hofstadter-Hubbard model captures the physics of strongly correlated electrons in an applied magnetic field, which is relevant to many recent experiments on Moir\'e materials. Few large-scale, numerically exact simulations exists for this model. In this work, we simulate the Hubbard-Hofstadter model using the determinant quantum Monte Carlo (DQMC) algorithm. We report the field and Hubbard interaction strength dependence of charge compressibility, fermion sign, local moment, magnetic structure factor, and specific heat. The gross structure of magnetic Bloch bands and band gaps determined by the non-interacting Hofstadter spectrum is preserved in the presence of $U$. Incompressible regions of the phase diagram have improved fermion sign. At half filling and intermediate and larger couplings, a strong orbital magnetic field delocalizes electrons and reduces the effect of Hubbard $U$ on thermodynamic properties of the system.
\end{abstract}

\begin{document}

\title{Thermodynamics of correlated electrons in a magnetic field}

\author{Jixun K. Ding}
\email{jxding@stanford.edu}
\affiliation{Department of Applied Physics, Stanford University, CA 94305, USA}

\author{Wen O. Wang}
\affiliation{Department of Applied Physics, Stanford University, CA 94305, USA}

\author{Brian Moritz}
\affiliation{Stanford Institute for Materials and Energy Sciences,
SLAC National Accelerator Laboratory, 2575 Sand Hill Road, Menlo Park, CA 94025, USA}

\author{Yoni Schattner}
\affiliation{Stanford Institute for Materials and Energy Sciences,
SLAC National Accelerator Laboratory, 2575 Sand Hill Road, Menlo Park, CA 94025, USA}
\affiliation{Department of Physics, Stanford University, Stanford, CA 94305, USA}

\author{Edwin W. Huang}
\affiliation{Department of Physics and Institute of Condensed Matter Theory, University of Illinois at Urbana-Champaign, Urbana, IL 61801, USA}

\author{Thomas P. Devereaux}
\email{tpd@stanford.edu}
\affiliation{Stanford Institute for Materials and Energy Sciences,
SLAC National Accelerator Laboratory, 2575 Sand Hill Road, Menlo Park, CA 94025, USA}
\affiliation{
Department of Materials Science and Engineering, Stanford University, Stanford, CA 94305, USA}
\date{\today}

\maketitle

\section{Introduction}

Strong magnetic fields allow us to probe the phase diagram of strongly correlated materials and uncover novel phases. For example, a magnetic field induces charge/pair density wave order in cuprate superconductors~\cite{Wu2011,Gerber2015,Jang2016, Edkins2019}; magnetic field is a convenient tuning parameter for accessing quantum critical points ~\cite{Grigera2001,Custers2003}; and field-induced reentrant superconductivity also has been reported in uranium compounds~\cite{Levy2005,Ran2019}. With the recent proliferation of experimental evidence for fractional quantum Hall effect, superconductivity, and other correlated electron phases~\cite{Hunt2013,Wang2015,Cao2018,Spanton2018,Yu2022,Saito2021} in graphene Moir\'e superlattices, there is renewed interest in studying the behavior of strongly correlated electronic systems in strong magnetic fields.

Properties of non-interacting electrons in a two-dimensional periodic lattice under the influence of a strong magnetic field are fairly well-understood. In this system, the competition between lattice and magnetic length scales leads to the fractal Hofstadter butterfly spectrum with recursive magnetic subband structure~\cite{Hofstadter1976,MacDonald1983}, which generalizes the idea of Landau levels in a free electron gas. The Chern numbers associated with these magnetic subbands provide an elegant explanation of the integer quantum Hall effect~\cite{Thouless1982}.
Experimentally, the Hofstadter Hamiltonian has been realized in ultra-cold atoms loaded on optical lattices~\cite{Miyake2013}, and direct observation of the Hofstadter spectrum has been reported in Moir\'e superlattices in graphene with high resolution~\cite{Dean2013,Ponomarenko2013}.

The most natural framework for understanding the simultaneous influence of magnetic field and Coulomb interaction on electrons in a periodic lattice is to take the Hofstadter Hamiltonian and add to it a Hubbard interaction term. In the literature, this is sometimes called the Hofstadter-Hubbard or Hubbard-Hofstadter model. This model has been investigated using Hartree-Fock mean-field theory~\cite{Gudmundsson1995,Doh1998}, exact diagonalization~\cite{Barelli1996,Czajka2006}, dynamical mean-field theory~\cite{Acheche2017,Markov2019}, and in the large $U$ limit via renormalized mean-field theory~\cite{Tu2018}. Aside from exact diagonalization, which is limited to small system sizes, all methods used to study the Hubbard-Hofstadter model have been approximate and don't capture the full extent of quantum fluctuations. It is not conclusive, for example, whether interactions change or preserve the gap structure of the Hofstadter butterfly~\cite{Doh1998,Czajka2006,Markov2019}.

Determinant quantum Monte Carlo (DQMC) \cite{Blankenbecler1981,Hirsch1985,White1989} is an unbiased and numerically exact algorithm for studying quantum systems at finite temperature. It employs a discrete Hubbard–Stratonovich transformation to reduce the quartic Hubbard interaction term to quadratic at the cost of introducing a fluctuating auxiliary field. This auxiliary field is then sampled using the Metropolis-Hastings algorithm. DQMC has been employed successfully in the (zero-field) Hubbard model to study spin and charge excitations~\cite{Jia2014,Kung2015} and  superconducting fluctuations~\cite{Khatami2015}, as well as find evidence for fluctuating stripes~\cite{Huang2017} and $T$-linear resistivity~\cite{Huang2019SM}. The DQMC method is especially powerful at half-filling in the absence of kinetic frustration~\cite{Varney2009}, where the fermion sign problem is absent due to particle-hole symmetry, even in the presence of a magnetic field. This allows simulations to be performed at much lower temperatures, providing access to properties more reflective of the ground state.

At half filling, the Fermi surface of the non-interacting Hofstadter model consists of a finite number of Dirac points at even-denominator rational fractions of magnetic flux per plaquette~\cite{Wen1989}.
The ground state of the Hubbard-Hofstadter model is thus expected to remain a Dirac semi-metal up to some finite coupling strength $U_{\mathrm{c}}$. At half a magnetic flux quantum per plaquette, the model also is known as the $\mathrm{\pi}$-flux model~\cite{Affleck1988}, and has been studied extensively numerically.
As interactions are turned on, the $\mathrm{\pi}$-flux model exhibits a quantum phase transition of the chiral Heisenberg Gross-Neveu universality class at $U_{\mathrm{c}} \approx 5.6t$ into an antiferromagnetic Mott insulator (AFMI)~\cite{Chang2012,Otsuka2014,Toldin2015,Otsuka2016,Guo2018}. 
Since the $\mathrm{\pi}$-flux model corresponds to the Hubbard-Hofstadter Hamiltonian threaded with maximum possible flux, we may think of the zero-field Hubbard model on a half-filled square lattice as the ``$0$-flux model'', which exibits a metal$-$AFMI transition with $U_{\mathrm{c}} = 0$~\cite{Hirsch1985,White1989}. Our simulations address intermediate field strengths between the $0$-flux and $\mathrm{\pi}$-flux Hubbard model, which, to the best of our knowledge, has not been studied via DQMC.

In this work, we study the Hubbard-Hofstadter model using DQMC and present the evolution of thermodynamic properties of correlated electrons in an orbital magnetic field $B$. We demonstrate that the gross structure of magnetic Bloch bands and band gaps determined by the non-interacting Hofstadter spectrum is preserved in the presence of $U$. Moreover, we determine that the many-body fermion sign is directly connected to electronic charge compressibility. Finally, focusing on the half-filled AFMI, we find that an orbital magnetic field tends to delocalize electrons and thus effectively lower the influence of Hubbard $U$.

\section{Methods}
We study the single-band Hubbard-Hofstadter model on a two dimensional square lattice
\begin{multline}
H = -t\sum_{ \langle ij\rangle \sigma}   \l\{\exp\left[\mathrm{i}\varphi_{ij}\right]c_{i \sigma}^\dagger c_{j \sigma} + \mathrm{h.c.}\r\}  \\
- \mu \sum_{i \sigma} n_{i\sigma} +U\sum_{i}\left(n_{i\uparrow} - 1/2 \right)\left(n_{i\downarrow} - 1/2\right), \label{eq:hamiltonian}
\end{multline}
where $t$ is the hopping integral between the nearest neighbor sites $\langle i j\rangle$, $\mu$ is chemical potential, and $U$ is the on-site Coulomb interaction strength. $c_{i\sigma}^{\dagger}$ ($c_{i\sigma}$) is the creation (annihilation) operator for an electron on site $i$ with spin $\sigma=\uparrow,\downarrow$ and $n_{i\sigma} =  c_{i\sigma}^\dagger c_{i\sigma}$ measures the number of electrons of spin $\sigma$ on site $i$. As this model only has nearest-neighbor hopping, it preserves particle-hole symmetry at half-filling with $\mu = 0$. A uniform, orbital magnetic field is introduced by the Peierls substitution via the phase
\begin{equation}
\varphi_{ij} = \dfrac{2\mathrm{\pi}}{\Phi_0} \int_{\vec{R}_i}^{\vec{R}_j} \vec{A}\cdot d\bm{\ell},
\end{equation}
where the integral is taken over the shortest straight line path, $\Phi_0 = h/e$ is the magnetic flux quantum, and $\mathbf{R}_i$ is the position of site $i$. We choose the symmetric gauge 
$\mathbf{A}= (-y\hat{x} + x\hat{y})B/2$ and do not include any Zeeman coupling terms.

We simulate the Hamiltonian in \cref{eq:hamiltonian} on a finite cluster with lattice constant $a=1$, and $N_x$ and $N_y$ sites in the $x$ and $y$ directions, respectively. $N = N_x N_y$ denotes the total number of sites. We implement modified periodic boundary conditions consistent with magnetic translation symmetry~\cite{Assaad2002}.
Requiring that the wave function be single-valued on the torus gives the flux quantization condition 
$\Phi/\Phi_0 = n_f/N$, where $\Phi=Ba^2$ is the flux through a plaquette and $n_f$ is an integer. 

Allowing the hopping integral to carry a complex phase requires us to modify the standard DQMC algorithm to use complex numbers, which increases the run-time of our algorithm $\sim 3$-fold. The complexified DQMC algorithm retains the same $O(M^3L)$ scaling as the real DQMC algorithm, where $M = N_x = N_y$ is the linear size of the lattice, and $L$ is the number of imaginary time discretization steps.
Unless otherwise specified, all Hubbard-Hofstadter DQMC simulations are performed on a $N_x = N_y = 8$ square cluster.  Error bars in DQMC results, when shown, denote $\pm 1$ standard error of the mean, estimated by jackknife resampling. Detailed simulation parameters are listed in Supplementary Note 1. Non-interacting results, where shown, are obtained from diagonalizing the Hofstadter model on a $40\times 40$ cluster in order to minimize finite size effects.

\section{Results and Discussion}

\subsection{Interacting Gap Structure}
\begin{figure*}
    \includegraphics[width=\textwidth]{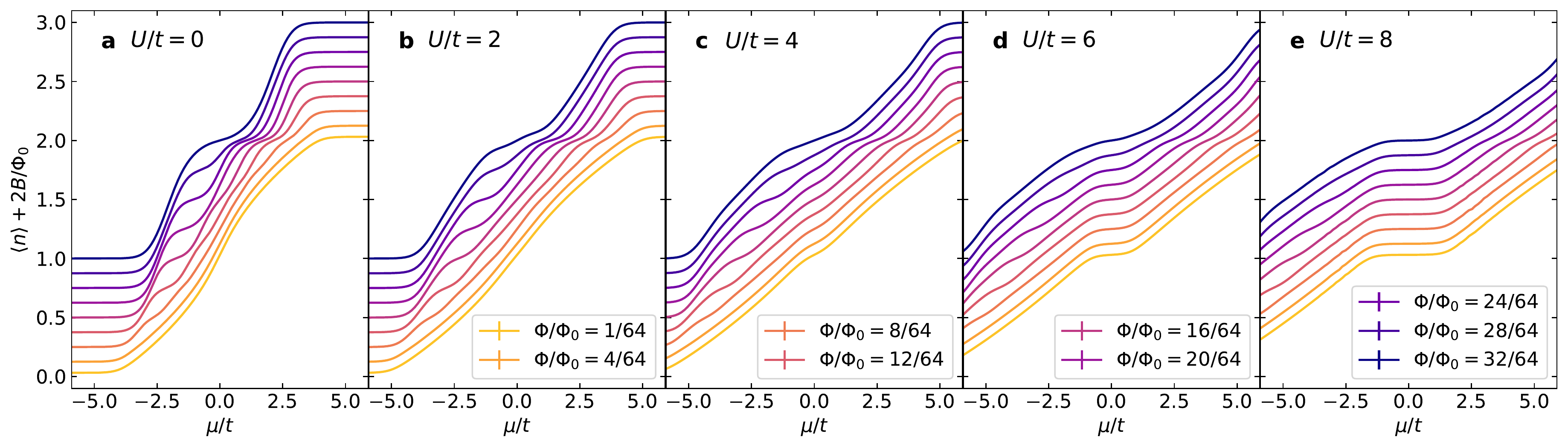} 
    \caption{Electron density $\langle n \rangle$ vs. chemical potential $\mu$ (a) in the non-interacting system, and (b)-(e) with Hubbard $U/t = 2-8$. Curves with the same color have the same magnetic field strength $\Phi/\Phi_0$ across all panels.Each curve is plotted with an offset $2B/\Phi_0$ in order to improve visibility of inflection points. Error bars, corresponding to $\pm 1$ standard error of the mean, estimated by jackknife resampling, are smaller than the size of data points.All plots have inverse temperature $\beta = 4/t$.}
    \label{density-vs-mu}
\end{figure*}

In \cref{density-vs-mu}, we show the electron density $\langle n\rangle$ vs. chemical potential $\mu$ at different field strengths for $U/t = 0-8$. In \cref{density-vs-mu}(a) we observe an electron density plateau where there are energy gaps between magnetic Bloch bands, $\langle n \rangle = 2B \nu/\Phi_0, \nu \in \mathbb{Z}$. 
As $U$ increases in \cref{density-vs-mu}(b)-(e), these plateaus are weakened and pushed outward in chemical potential, but inflections of the $\langle n\rangle$ vs. $\mu$ curves are still visible at the same density values, indicating that degeneracy of Landau levels is not modified by $U$. Since we are at relatively high temperature, only the most prominent band gaps (with Chern number $C=\pm 1$) $\langle n \rangle = 2B/\Phi_0$ remain visible at larger $U$ values. 
Additionally, at half filling, as $U$ increases, a Mott gap  appears and widens for all values of magnetic field. 
In \cref{density-vs-mu}(d)-(e), for $U/t \geq 6$, when the Mott gap is well-defined, it decreases monotonically as the magnetic field increases, consistent with previous exact diagonalization results \cite{Czajka2006}. The same trend can be seen in a ``correlated Hofstadter butterfly'' plot, as shown in Supplementary Fig. S1 and described in Supplementary Note 2. We will discuss later, and in more detail, the behavior of the Mott gap.

\begin{figure}
    \includegraphics[width=\linewidth]{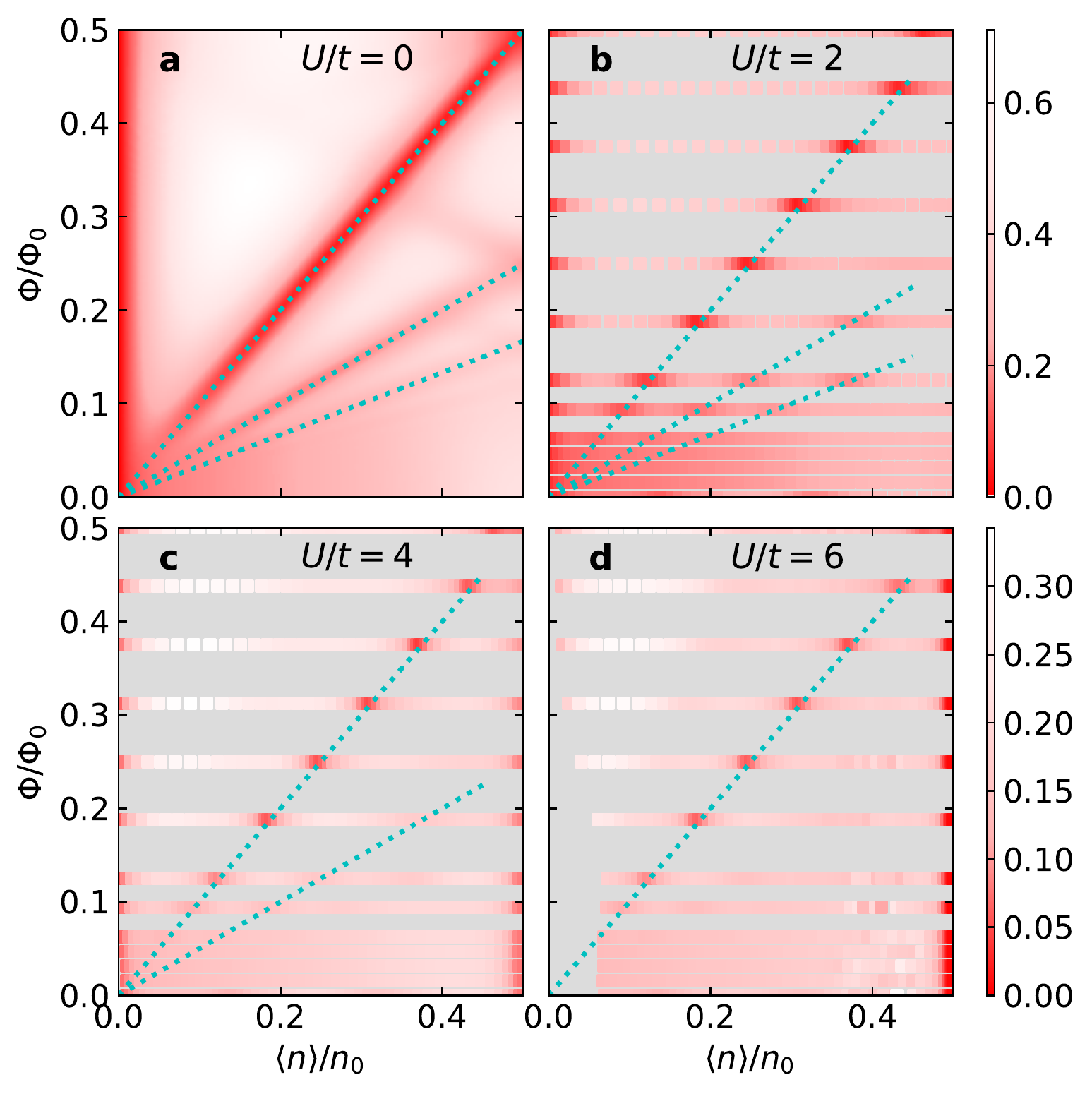} 
    \caption{Wannier diagrams (a) in the non-interacting system and (b)-(d) with Hubbard $U/t = 2-6$. Grey regions in (b)-(d) are parameter regions where we don't have simulation data. All plots have inverse temperature $\beta = 5/t$. Magnetic field strength is displayed as $\Phi/\Phi_0$, and electron density is shown as $\langle n \rangle /n_0$, where $n_0$ is the electron density of a completely filled system, which in our case is $2$. The system is particle-hole symmetric, so only the range $\langle n \rangle /n_0 \in [0,0.5]$ is shown. Dotted cyan lines indicate where we expect local minima of charge compressibility to occur from the non-interacting model. }
    \label{wannier-plot}
\end{figure}

It is instructive to plot our data as Wannier diagrams~\cite{Wannier1978}, i.e.~color intensity plots of charge compressibility $\chi = \partial\langle n\rangle/\partial\mu$ as a function of electron density $\langle n \rangle$ and magnetic field strength $B$. Charge compressibility, or thermodynamic density of states, is directly measurable in experiments \cite{Hunt2013,Yu2022}. In a non-interacting system, at zero temperature, charge compressibility is equivalent to the single-particle density of states. We measure charge compressibility in DQMC simulations as
\begin{equation}
\chi = \frac{\beta}{N} \sum_{ij}\l[\langle n_i n_j\rangle  - \langle n_i \rangle \langle n_j \rangle \r],
\end{equation}
where $n_i = n_{i\uparrow} + n_{i\downarrow}$. 
In \cref{wannier-plot}, we show Wannier diagrams for $U/t = 0-6$. For all values of $U$, we observe local minima of $\chi$ (indicating incompresssible states) along straight lines satisfying the Diophantine equation
\begin{equation}
\frac{\langle n\rangle }{n_0} = r\l(\frac{\Phi}{\Phi_0}\r) + s, \label{diophantine}
\end{equation}
where $r$ and $s$ are integers, and $n_0=2$ is the electron density of the completely filled system. This is consistent with what we expect from the Hofstadter spectrum in the non-interacting system~\cite{Wannier1978}.
The most prominent incompressible state with $r = 1, s = 0$ remains clearly visible up to $U /t= 8$. Less prominent incompressible states with $r = 2$ and $r=3$ persist to $U/t=4$ and $U/t=2$, respectively. These results show that the integer quantum Hall states for $r \leq 3, s= 0$ have no weak coupling instabilities with respect to Hubbard repulsion; the $r = 1, s = 0$ state remains stable up to large $U$.
At half filling, the vertical compressibility minima indicative of the Mott gap becomes visible for $U/t \gtrsim 4$. Thus, we argue that the gross structure of magnetic Bloch bands and band gaps determined by the non-interacting Hofstadter spectrum is preserved in the presence of $U$, with the Mott gap at half-filling when $U/t \gtrsim 4$ superimposed as an additional feature. Due to the sign problem, our DQMC simulations are restricted to relatively high temperature $\beta t \leq 5$, so we cannot resolve conclusively how much $U$ changes the fine structure of the Hofstadter spectrum.

\subsection{Fermion Sign}

An important quantity in QMC simulations of interacting fermions is the fermion sign. The Hubbard-Hofstadter model is sign-problem-free at half filling on a bipartite lattice. But the fermion sign problem~\cite{Loh1990} fundamentally prevents us from obtaining high quality simulation data at low temperatures and away from half filling. Thus, any insight into factors affecting the severity of the sign problem is valuable. 
Since the fermion sign problem is NP-hard~\cite{Troyer2005}, we do not expect a general solution to the fermion sign problem to exist. Nevertheless, as the sign problem is representation-dependent, it is possible to reduce or completely remove the sign problem for specific classes of non-generic Hamiltonians \cite{Li2019}. 

\begin{figure}
    \includegraphics[width=0.9\linewidth]{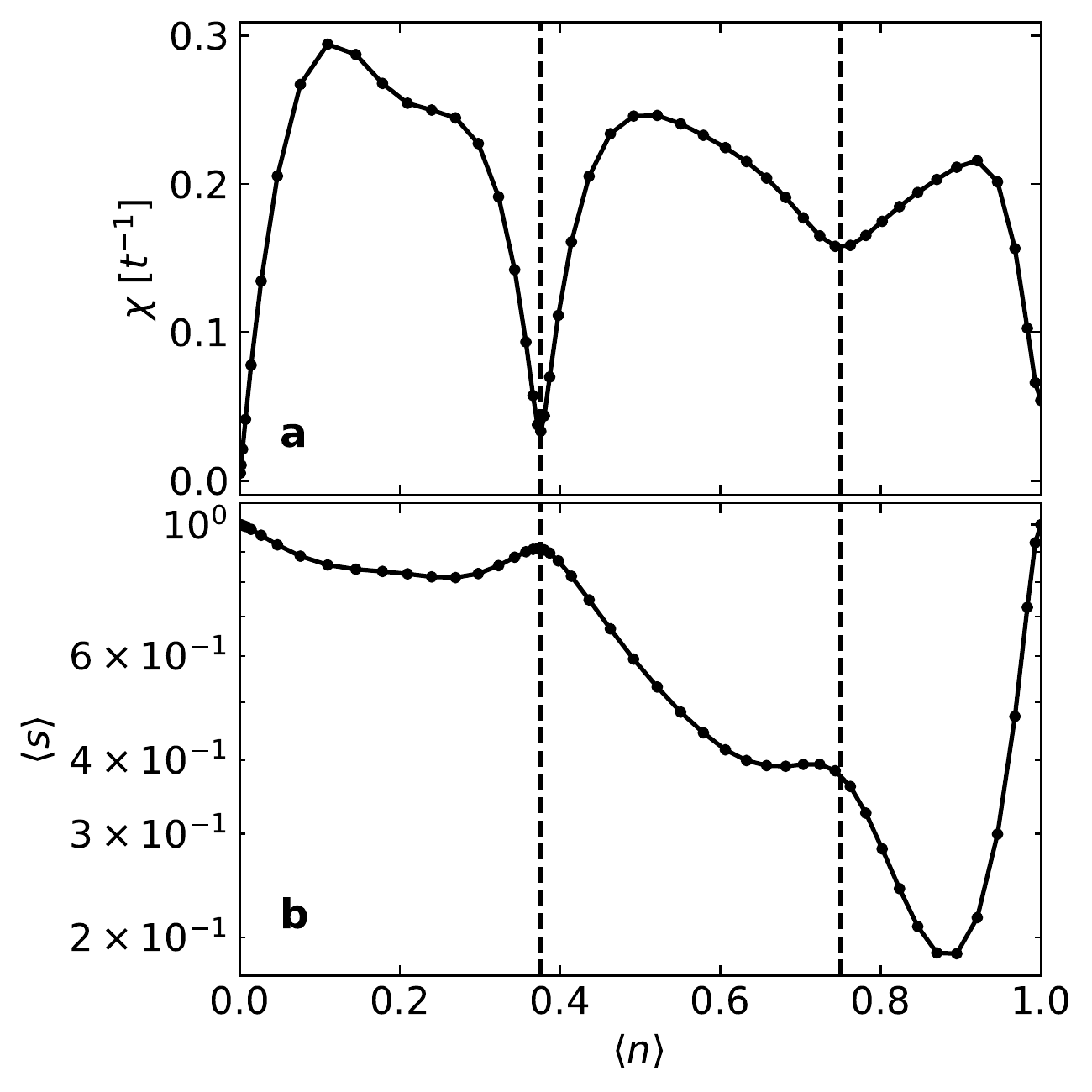} 
    \caption{(a) Charge compressibility $\chi =  \partial\langle n \rangle/\partial \mu$ and (b) fermion sign $\langle s \rangle$ plotted against electron density $\langle n \rangle$ at magnetic field strength $\Phi/\Phi_0 = 12/64$, inverse temperature $\beta = 6/t$ and Hubbard interaction $U/t=4$. This system is particle-hole symmetric, so we only plot the density range $\langle n \rangle \in [0,1]$. Dashed lines indicate electron densities at which $\chi$ reaches local minina and $\langle s \rangle$ reaches local maxima. Error bars, corresponding to $\pm 1$ standard error of the mean, estimated by jackknife resampling, are smaller than the size of data points.}
    \label{compress-sign-special}
\end{figure}

In this work, we find a correlation between the fermion sign and the charge compressibility. 
In \cref{compress-sign-special}, we show the fermion sign $\langle s \rangle$ and charge compressibility $\chi$, both plotted against $\langle n \rangle$, for one representative set of parameters.   Local minima of charge compressibility in this interacting system exactly correspond to local maxima of the fermion sign. At these local maxima, the fermion sign may be an order of magnitude improved over its value at other electron densities and that of the standard zero-field Hubbard model. For an extended figure demonstrating that this correspondence is general across our parameter space and not a finite size artifact, see Supplementary Fig. S2 and Supplementary Note 3.
Our results may mean that although the Hubbard model in general suffers from a sign problem, it is possible to obtain good results when we are precisely located on an integer quantum Hall plateau. Since similar sign-compressibility correspondence has been reported~\cite{White1989,Mondaini2012,Kung2016,Huang2019}, it appears that the improvement of fermion sign in insulating phases is quite general, consistent with our intuition that fermionic statistics become less important in localized states.
Our results also relate to recent work~\cite{Mondaini2022-pub,Wessel2017,Goetz2022-pub} suggesting that the fermion sign is not merely a coincidental barrier to accessing low-temperature physics, but may be reflective of intrinsic physics of model Hamiltonians. 

\subsection{Half-Filling}

Finally, we focus on half filling, where we believe interesting interplay between Hofstadter physics and Hubbard physics occurs.  In the absence of a magnetic field, the ground state of the half-filled Hubbard model is an AFMI at any nonzero value of $U$, i.e. ($U_{\mathrm{c}} = 0$)~\cite{Hirsch1985,White1989}, due to perfect nesting of the Fermi surface and a logarithmically divergent single-particle density of states. In the limit of strong interactions $U\gg t$, the half-filled Hubbard model maps to the Heisenberg model with antiferromagnetic nearest neighbor spin exchange energy $J = 4t^2/U$~\cite{Fazekas1999}. 

In the presence of an orbital magnetic field, the non-interacting density of states at half filling is modified significantly. As can be seen in \cref{wannier-plot}(a), the density of states/charge compressibility does not change monotonically with field, but instead shows prominent minima at $\Phi/\Phi_0 = p/q$, where $p$ and $q$ are co-prime and $q$ is even, corresponding to a non-interacting ground state with $q$ inequivalent Dirac cones \cite{Wen1989}. The large-field limit corresponds to the $\mathrm{\pi}$-flux model, in which a semimetal$-$AFMI transition occurs at $U_{\mathrm{c}} \approx 5.6 t$. 
As the orbital magnetic field significantly changes the non-interacting density of states at half filling, we expect that critical $U_{\mathrm{c}}$ should exhibit $B$ field dependence. 
It would be interesting to investigate if $U_{\mathrm{c}}$ changes monotonically with $B$, or if it exhibits non-monotonicity commensurate with the oscillatory behavior of the density of states. We defer the mapping of this $U$-$B$ phase diagram to future work. 

For the remainder of this section, we focus on the parameter region $U/t \in [6,10]$. Here, the system is safely an AFMI at all field strengths. We examine the evolution of local magnetic moment $\langle m_z^2 \rangle$, antiferromagnetic structure factor $S(\mathrm{\pi},\mathrm{\pi})$ and specific heat $c_v = \partial\langle E\rangle /\partial T$ with magnetic field strength, and see that these thermodynamic quantities all consistently show that a strong orbital magnetic field tends to modify the AFMI by delocalizing electrons and thereby reducing the effect of $U$ on the low-energy properties of the Mott insulating phase. For finite-size analysis of thermodynamic observables at half filling, see Supplementary Fig. S3 and Supplementary Note 3.

\begin{figure}
    \includegraphics[width=0.95\linewidth]{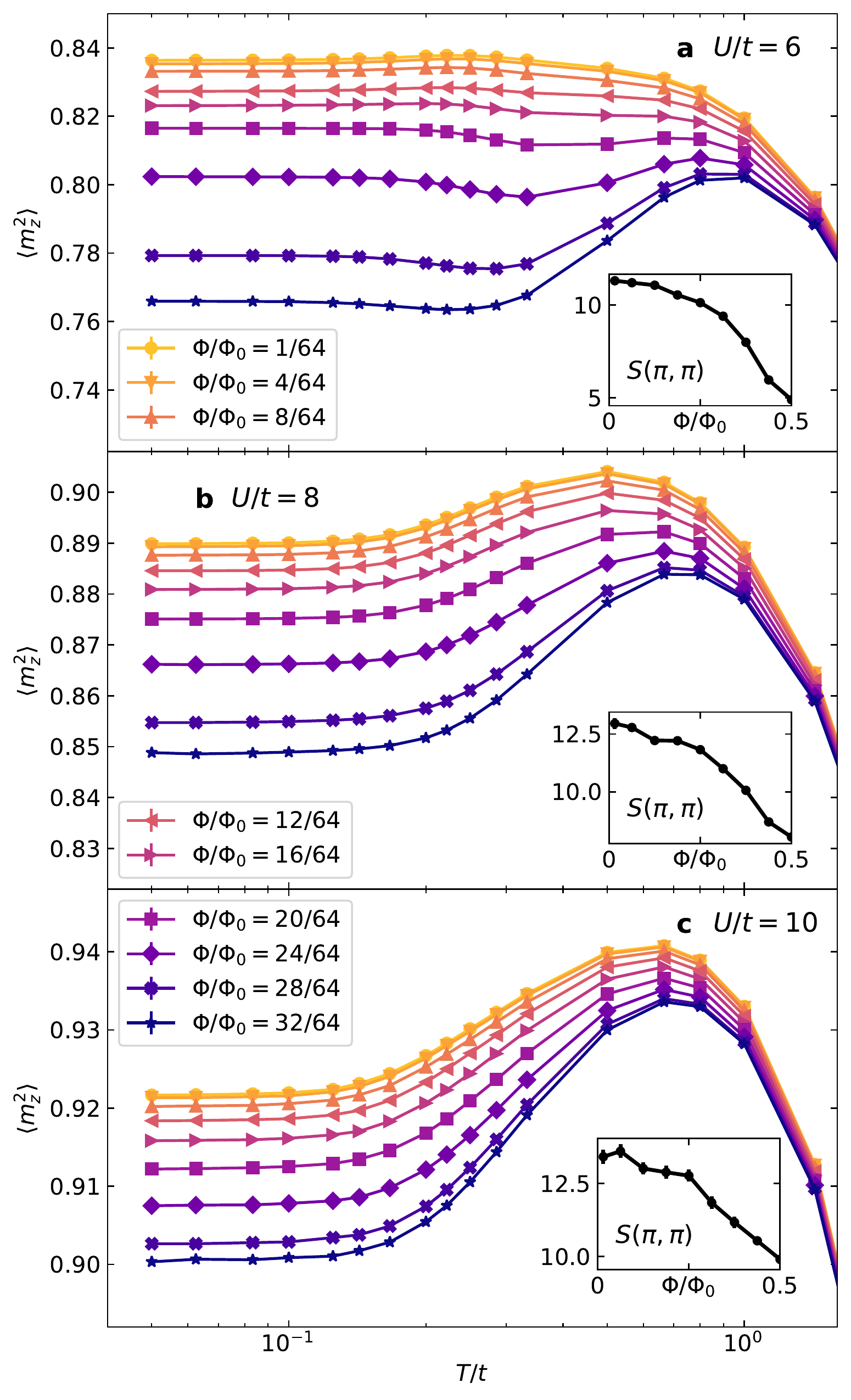} 
    \caption{Temperature and field dependence of local moment $\langle m_z^2\rangle$ at half filling. Insets display the field dependence of magnetic structure factor $S(\mathrm{\pi},\mathrm{\pi})$ at $\beta t = 16$ for magnetic field strength $\Phi/\Phi_0 \in [0,0.5]$. (a)-(c) Correspond to Hubbard interaction strength $U/t=6-10$, respectively. Curves with the same color and marker type have the same magnetic field strength across all panels. Error bars, corresponding to $\pm 1$ standard error of the mean, estimated by jackknife resampling, are smaller than the size of data points.}
    \label{mz-half-fill-all}
\end{figure}

\begin{figure}
    \includegraphics[width=\linewidth]{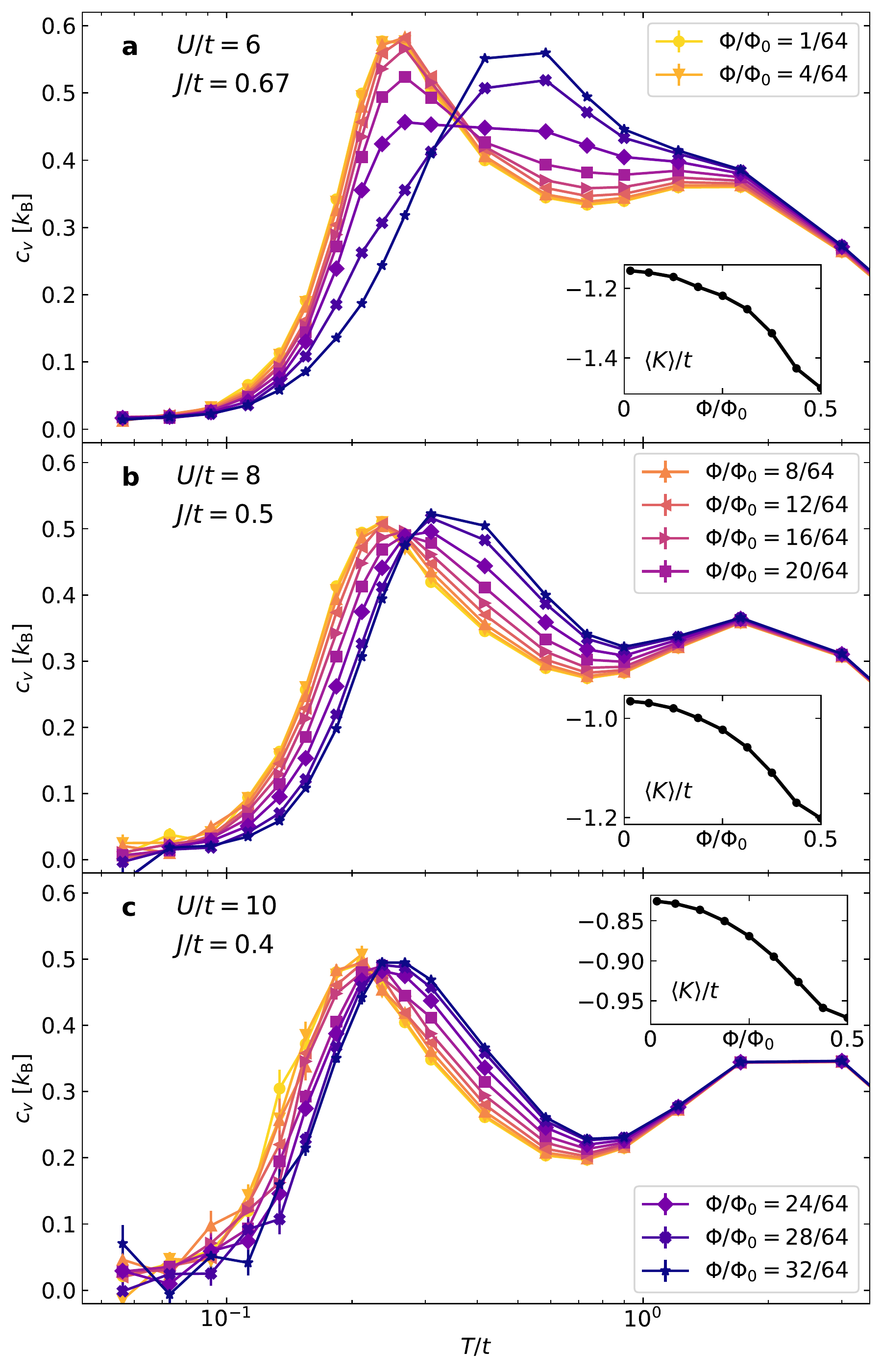} 
    \caption{Temperature and field dependence of specific heat $c_v$ at half filling. Insets display the field dependence of kinetic energy $\langle K \rangle$ at $\beta t = 16$ for magnetic field strength $\Phi/\Phi_0 \in [0,0.5]$. (a)-(c) Correspond to Hubbard interaction strength $U/t=6-10$, respectively. Values of $J=4t^2/U$ are also indicated for each $U$. Curves with the same color and marker type have the same magnetic field strength across all panels. Error bars denote $\pm 1$ standard error of the mean, estimated by jackknife resampling.}
    \label{cv-half-fill}
\end{figure}

In \cref{mz-half-fill-all}, we show the temperature and field dependence of the local moment. The local moment or sublattice magnetization
\begin{equation}
\langle m_z^2\rangle = \frac{1}{N}\sum_i \langle (n_{i\uparrow} - n_{i\downarrow})^2\rangle
\end{equation}
measures the degree of spin localization. It is $0.5$ in the non-interacting system and approaches $1$ in the $U/t\rightarrow \infty$ limit. In the zero-field Hubbard model, $\langle m_z^2\rangle$ has features at $T\sim U$ associated with the formation of local moments, and at $T\sim J$, associated with short- or long-range ordering of local moments~\cite{Paiva2001,Varney2009}. We see that at fixed $U$, increasing magnetic field strength reduces the local moment monotonically at all temperatures, with the effect largest below temperatures $T\sim J$. 
The zero-field and $\mathrm{\pi}$-flux limit of local moment data are consistent with previous work~\cite{Guo2018}.

The magnetic structure factor is the Fourier transform of the real-space spin-spin correlation function
\begin{equation}
S(\vec{Q}) = \frac{1}{N}\sum_{ij} e^{\mathrm{i}\vec{Q}\cdot (\vec{R}_i-\vec{R}_j)}\langle (n_{i\uparrow}-n_{i\downarrow}) (n_{j\uparrow}-n_{j\downarrow})\rangle.
\end{equation}
When the system has long-range antiferromagnetic order, the structure factor is strongly peaked at the ordering wave vector $\vec{Q} = (\mathrm{\pi},\mathrm{\pi})$, with peak height scaling linearly with lattice size~\cite{Varney2009,Huse1988}. In our simulations, we find that at all $B$ and $U$ values, the magnetic structure factor is sharply peaked at $(\mathrm{\pi},\mathrm{\pi})$, consistent with the system being in the AFMI phase. Insets to \cref{mz-half-fill-all} show that at all $U$, the magnetic field monotonically reduces $S(\mathrm{\pi},\mathrm{\pi})$, indicating that the magnetic field reduces the AFMI ordering tendencies.

\cref{cv-half-fill} shows the evolution of the low temperature peak of $c_v$ with magnetic field and Hubbard $U$.
We calculate $c_v$ numerically by measuring energy as a function of temperature $\langle E(T)\rangle$ and taking the finite difference $\Delta \langle E\rangle / \Delta T$. In the zero-field half-filled Hubbard model, at large $U$, the specific heat has a ``two peak'' structure, with a broad high temperature peak at $T \sim U$  associated with charge fluctuations and and a narrow low temperature peak at $T \sim J$ associated with spin fluctuations~\cite{Paiva2001,Wang2021}.
When a magnetic field is turned on, as shown in \cref{cv-half-fill}(b)-(c), $U$ is large enough that the two peaks remain well-separated. The high temperature peak doesn't move, while the low temperature peak shifts to higher temperatures. In \cref{cv-half-fill}(a), the two peaks are initially well-separated at low fields, but at $\Phi/\Phi_0 \gtrsim 1/4$, the low temperature peak shifts upwards and merges with the high-temperature peak, complicating our interpretation. This is likely due to $U/t = 6$ being low enough for the system to not simply map to the Heisenberg model, and for the system to be close to the AFMI phase transition at $\mathrm{\pi}$-flux.

Since the low temperature peak in specific heat is associated with the spin exchange energy $J$, we are tempted to say that the orbital magnetic field increases $J$. However, this interpretation may be overly naive. We believe the more accurate statement is that the orbital magnetic field tends to delocalize electrons, and thus, effectively lower the influence of $U$ on low energy properties of the system. Insets to \cref{cv-half-fill} show that a magnetic field increases (in magnitude) kinetic energy in the insulating phase, which supports this interpretation.
This runs contrary to our usual intuition that an orbital magnetic field localizes electrons by winding them up into Landau orbits. However, here our starting point is a correlated insulator, rather than free electrons (or a Fermi liquid). Our results in \cref{mz-half-fill-all,cv-half-fill}, along with the decreasing width of the Mott gap in \cref{density-vs-mu}(d)-(e), suggest that in the AFMI phase, the orbital magnetic field tends to delocalize electrons, increase kinetic energy, and lower the effective influence of $U$. We observe that the influence of magnetic field is suppressed as $U$ increases. As $U/t \rightarrow \infty$, the influence from the $B$ field will diminish and become negligible, since in the atomic limit, no hopping exists, and the orbital magnetic field cannot have an influence on the system.

\section{Conclusions}
In this work, we implemented DQMC to simulate the Hubbard-Hofstadter model and directly investigate field dependent thermodynamic properties of correlated electrons, specifically focusing on the charge compressibility, local moment, magnetic structure factor, and specific heat. 
By examining charge compressibility, we find that magnetic Bloch bands are smeared out by the Hubbard interaction, but the non-interacting band gaps away from half-filling persist at temperatures accessible to DQMC in the presence of local correlations. 
At half filling, we find that the orbital magnetic field and Hubbard potential act antagonistically. At intermediate to strong coupling $U/t \in [6,10]$, strong orbital magnetic fields reduce the apparent width of the Mott gap, reduce the magnitude of the local moment and magnetic structure factor, increase kinetic energy, and shift the low-$T$ peak of specific heat to higher temperatures. Together, these phenomena indicate that an orbital magnetic field tends to delocalize electrons and reduce the effect of $U$. From the algorithmic perspective, we find that the fermion sign in DQMC simulations is improved significantly when the physical system is incompressible.

For future work, we are interested in mapping out the $U$-$B$ phase diagram for $U/t \in [0,6]$ and in studying the evolution of $U_{\mathrm{c}}$ with magnetic field strength. At smaller values of $U$, we can achieve lower temperatures in DQMC simulations, but finite size effects also become more significant. Pinning down the precise location of the phase transition will require simulations on much larger lattices, as well as careful finite size scaling analysis.

As this work is partially motivated by understanding the large variety of exotic states in Moir\'e materials~\cite{Hunt2013,Wang2015,Spanton2018,Cao2018,Yu2022,Saito2021},
another potential direction for future work is to use methods developed here to study the Hubbard-Hofstadter model on the honeycomb or triangular lattice. It has been suggested that the single-band Hubbard model on a triangular lattice is directly applicable to transition metal dichalcogenide heterobilayers~\cite{Wu2018,Tang2020}. However, due to the linear dispersion relation of Dirac fermions, the Coulomb interaction is poorly screened in graphene. It is likely that some multi-orbital, extended Hubbard model may be required to capture the full effect of electron correlations, for example, in twisted bilayer graphene~\cite{Andrews2020,Po2019}. 


\section{Data Availability}
Aggregated numerical data and analysis routines required to reproduce the figures can be found at \url{https://doi.org/10.5281/zenodo.6383764}. Raw simulation data that support the findings of this study are stored on the Sherlock cluster at Stanford University and are available from the corresponding author upon reasonable request.

\section{Code Availability}
The most up-to-date version of our DQMC simulation code can be accessed at \url{https://github.com/edwnh/dqmc}.

\section{Author Contributions}
EWH wrote the DQMC simulation code. JKD performed simulations and data analysis. TPD conceptualized the work. All authors (JKD, WOW, BM, YS, EWH, TPD) participated in discussions and manuscript writing.

\section{Competing Interests}
The authors declare no competing interests.

\section{Acknowledgements}
We acknowledge helpful discussions with Richard Scalettar, Philip W. Phillips, Allan.H. MacDonald, Benjamin E. Feldman,
Young S. Lee and Jiachen Yu.

This work was supported by the U.S. Department of Energy (DOE), Office of Basic Energy Sciences,
Division of Materials Sciences and Engineering. 
EWH was supported by the Gordon and Betty Moore Foundation's EPiQS Initiative through grants GBMF 4305 and GBMF 8691. YS was supported by the Gordon and Betty Moore Foundation’s EPiQS Initiative through grants GBMF 4302 and GBMF 8686.
Computational work was performed on the Sherlock cluster at Stanford University and on resources of the National Energy Research Scientific Computing Center, supported by the U.S. DOE, Office of Science, under Contract no. DE-AC02-05CH11231.

\bibliography{main}

\end{document}


\title{Supplementary Information: Thermodynamics of correlated electrons in a magnetic field}

\author{Jixun K. Ding}
\email{jxding@stanford.edu}
\affiliation{Department of Applied Physics, Stanford University, CA 94305, USA}

\author{Wen O. Wang}
\affiliation{Department of Applied Physics, Stanford University, CA 94305, USA}

\author{Brian Moritz}
\affiliation{Stanford Institute for Materials and Energy Sciences,
SLAC National Accelerator Laboratory, 2575 Sand Hill Road, Menlo Park, CA 94025, USA}

\author{Yoni Schattner}
\affiliation{Stanford Institute for Materials and Energy Sciences,
SLAC National Accelerator Laboratory, 2575 Sand Hill Road, Menlo Park, CA 94025, USA}
\affiliation{Department of Physics, Stanford University, Stanford, CA 94305, USA}

\author{Edwin W. Huang}
\affiliation{Department of Physics and Institute of Condensed Matter Theory, University of Illinois at Urbana-Champaign, Urbana, IL 61801, USA}

\author{Thomas P. Devereaux}
\email{tpd@stanford.edu}
\affiliation{Stanford Institute for Materials and Energy Sciences,
SLAC National Accelerator Laboratory, 2575 Sand Hill Road, Menlo Park, CA 94025, USA}
\affiliation{
Department of Materials Science and Engineering, Stanford University, Stanford, CA 94305, USA}
\date{\today}

\maketitle

\section{Supplementary Note 1: Simulation Parameters}
\label{sec:sim-params}

Determinant quantum Monte Carlo (DQMC) data at general filling levels shown in main text Figs. 1-3 are obtained from simulations performed using $4\times 10^3$ to $5\times 10^3$ warm-up sweeps  and $2.4\times 10^4$ to $1.05 \times 10^5$ measurement sweeps through the auxillary field. We run 10 independently seeded Markov chains for each set of parameters. For all parameter values, the imaginary time discretization interval $\Delta\tau \leq 0.1/t$, and the number of imaginary time slices $L = \beta/\Delta \tau \geq 10$.

Half-filling data shown in main text Figs. 4 and 5  are obtained from simulations performed with $2 \times 10^4$ to $4\times 10^4$ warm-up sweeps and $2 \times 10^5$ to $5 \times 10^5$ measurement sweeps. We run $20$ to $200$ independently seeded Markov chains for each set of parameters. At half filling, we enforce a smaller imaginary time discretization interval in order to reduce effects from Trotter error. Here the imaginary time discretization interval $\Delta\tau \leq 0.05/t$, and the number of imaginary time slices $L = \beta/\Delta \tau \geq 10$. We check that $\Delta \tau = 0.05/t$ is sufficiently small that Trotter error is negligible in thermodynamic observables and do not affect the conclusions presented in this paper.

In all simulations, multiple equal-time measurements are taken in each full measurement sweep through the auxillary field. Each Markov chain with $M$  measurement sweeps collects $M L /5$ equal-time measurements. The mean and standard error of observables are estimated via jackknife resampling of independent Markov chains.

\section{Supplementary Note 2: Correlated Hofstadter Butterfly}
\label{sec:correlated-butterfly}

\begin{figure}[htb]
    \includegraphics[width=\linewidth]{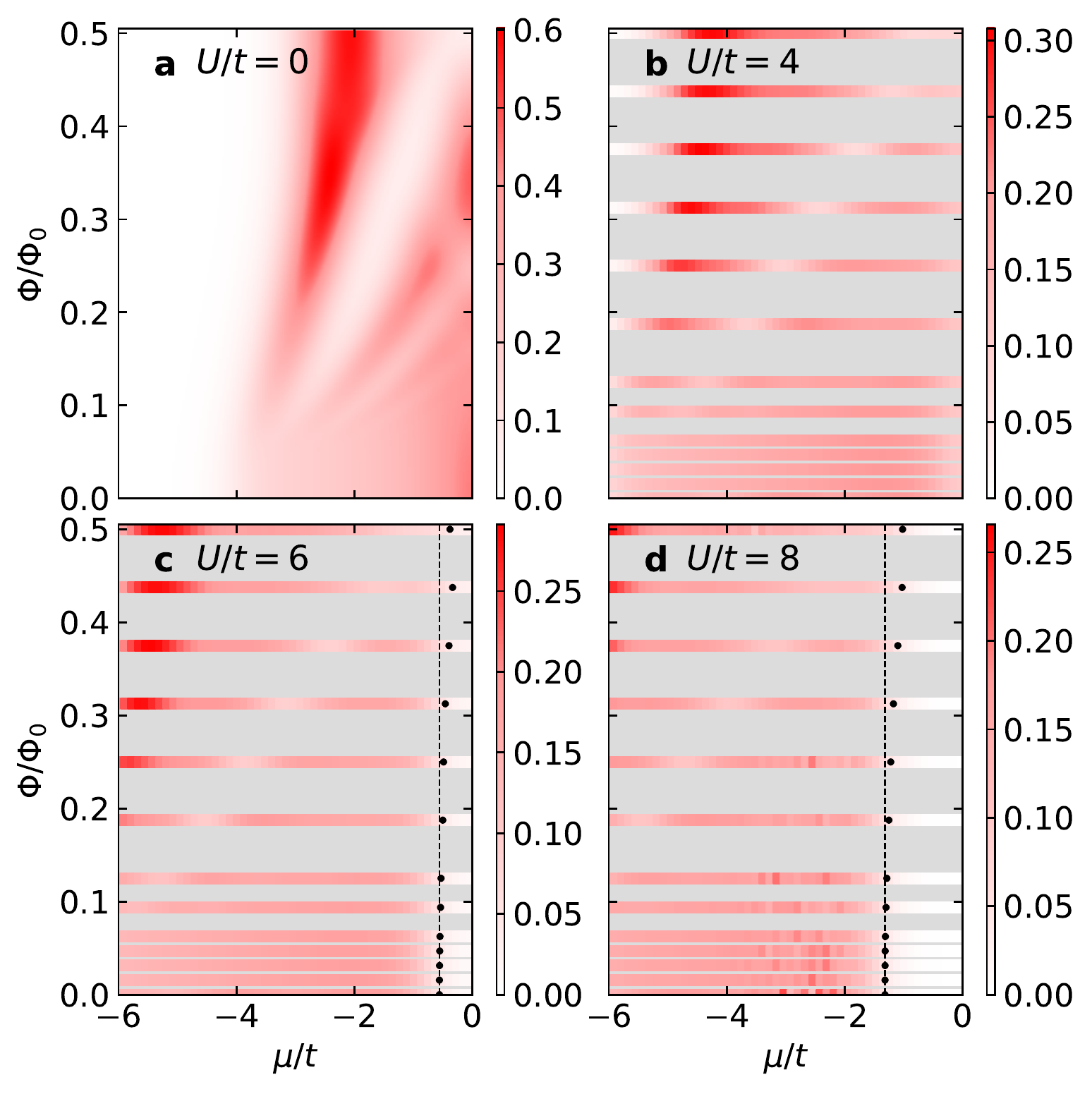}
    \caption{Color intensity plots of charge compressibility (a) in the non-interacting system and (b)-(d) with Hubbard $U/t = 4-8$. Grey regions in (b)-(d) are parameter regions where we don't have simulation data. 
 For (c) $U/t = 6$ and (d) $U/t=8$, black dots denote the approximate boundary of the Mott gap, determined by performing cubic spline interpolation on the $\chi(\mu)$ data, and solving for the chemical potential at which $\chi = 0.05/t$. Black dashed vertical lines mark the zero-field Mott gap boundary, demonstrating that the Mott gap shrinks as orbital magnetic field increases.  All subplots have inverse temperature $\beta = 4/t$. 
Magnetic field strength is displayed as $\Phi/\Phi_0$, and electron density is shown as $\langle n \rangle /n_0$, where $n_0$ is the electron density of a completely filled system, which in our case is $2$. The system is particle-hole symmetric, so only the range $\langle n \rangle /n_0 \in [0,0.5]$ is shown. }
    \label{correlated-butterfly}
\end{figure}

This section shows the ``correlated Hofstadter butterfly'' plot, which is a color intensity plot of charge compressibility $\chi = \partial\langle n\rangle/\partial\mu$ as a function of chemical potential and magnetic field strength $B$. Charge compressibility is treated as a proxy for local density of states in the correlated system, since we don't have enough data to perform Maximum Entropy analytic continuation to obtain the ``true'' local density of states. In \cref{correlated-butterfly}(c)-(d), we see the width of the apparent Mott gap shrinks as field strength increases. This is consistent with main text Fig.1(d)-(e) and our understanding of the effect of field on the Mott insulator in the half-filled Hubbard-Hofstadter model.

\begin{figure}
   \includegraphics[width=\linewidth]{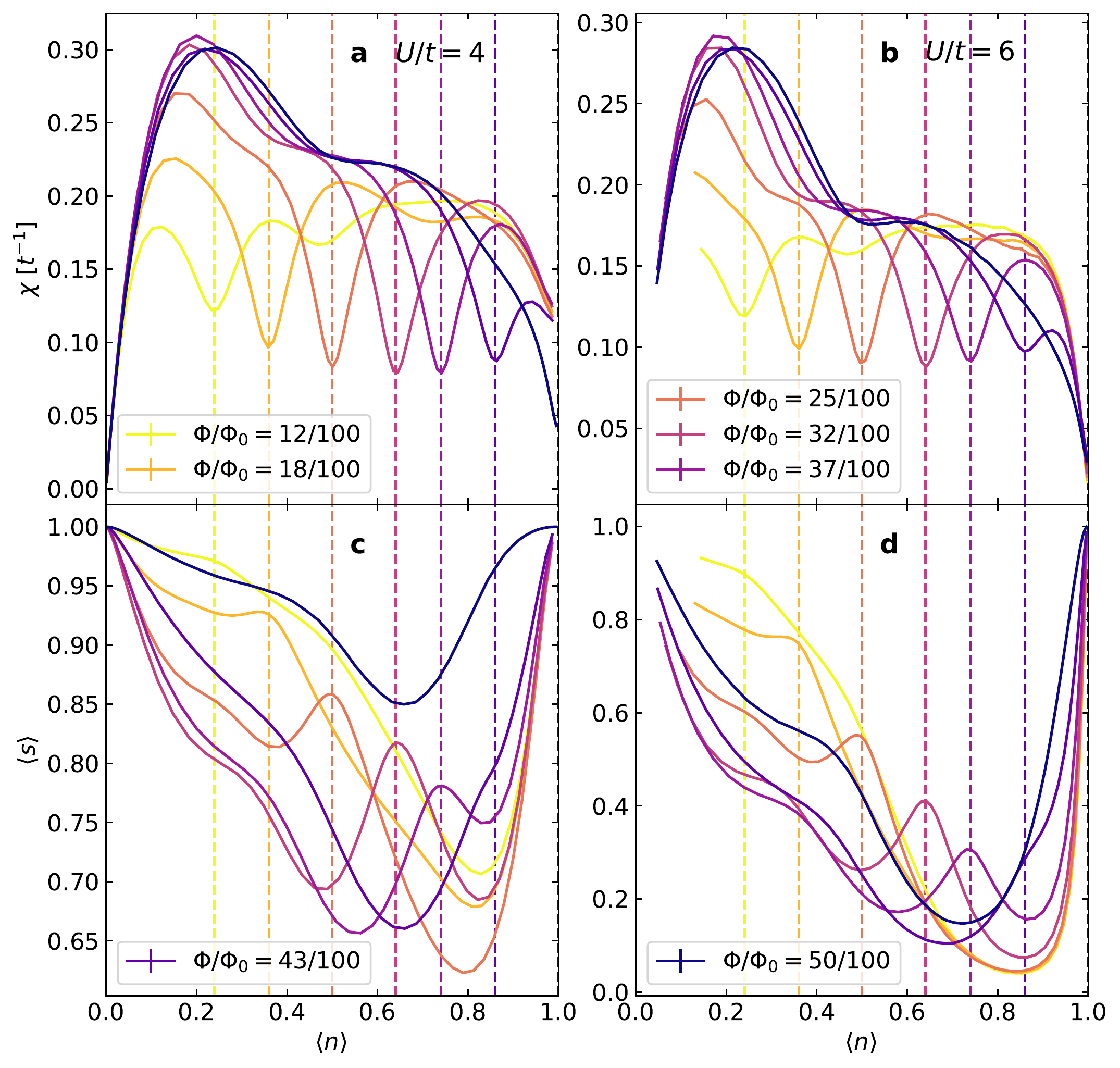} 
    \caption{(a)-(b) charge compressibility $\chi =  \partial\langle n \rangle/\partial \mu$ and (c)-(d) fermion sign $\langle s \rangle$ plotted against electron density $\langle n \rangle$. Results are obtained on $10\times 10$ lattice, with inverse temperature $\beta = 4/t$, and Hubbard interaction strengths $U/t=4-6$. This system is particle-hole symmetric, so we only plot the density range $\langle n \rangle \in [0,1]$. Curves with the same color have the same magnetic field strength $\Phi/\Phi_0$ across all panels. Vertical dashed lines are color-coded according to magnetic field strength, and indicate electron densities at which the charge compressibility reaches its first local minina, and where the fermion sign may have a corresponding local maxima or inflection point. Magnetic field strength is displayed as $\Phi/\Phi_0$. Error bars ($\pm 1$ standard error of the mean, estimated by jackknife resampling) are smaller than the size of data points.}
    \label{compress-sign-general}
\end{figure}

\section{Supplementary Note 3: Finite size analysis}
\label{sec:finite-size}

Since our DQMC simulations are limited to relatively small clusters, it's important to check if the conclusions of this paper correctly extrapolate to the thermodynamic limit. Extended data plot \cref{compress-sign-general}, obtained on a $10 \times 10$ lattice, shows that charge compressibility minima and fermion sign maxima occur at electron filling factors $\langle n \rangle = 2B \nu/\Phi_0, \nu \in \mathbb{Z}$, regardless of lattice size. The sign-compressibility correspondence is general across field strengths and Hubbard interactions and is not a finite-size artifact. 

\begin{figure}[b!]
    \includegraphics[width=0.95\linewidth]{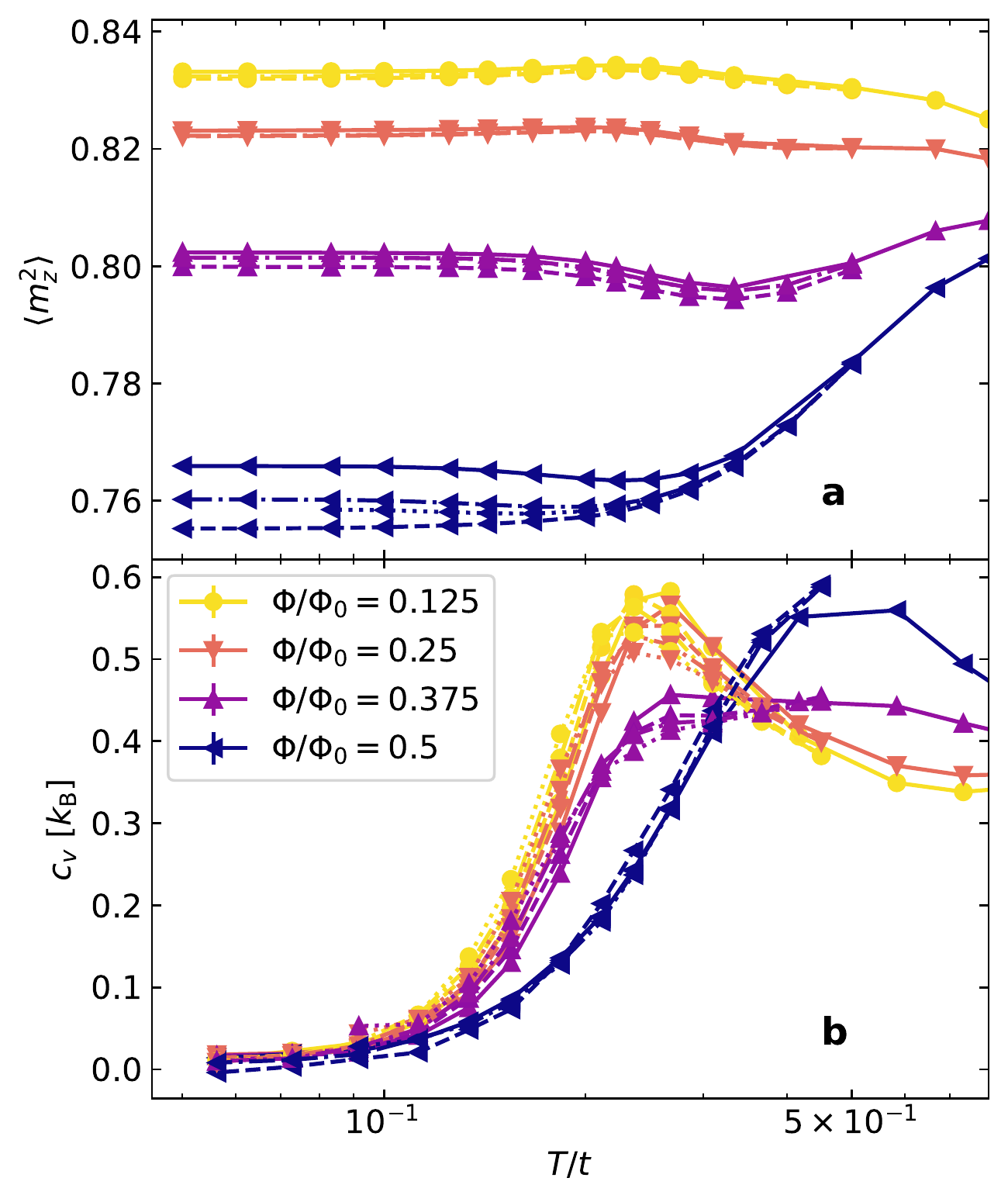} 
    \caption{Lattice size dependence of (a) local moment $\langle m_z^2\rangle$ and (b) specific heat $c_v$. Hubbard interaction strength $U/t = 6$. Solid, dashed, dot-dash, and dotted lines denote data obtained on $8\times 8$, $10\times 10$, $12\times 12$, and $16\times 16$ lattices, respectively. Magnetic field strength is displayed as $\Phi/\Phi_0$. Error bars ($\pm 1$ standard error of the mean, estimated by jackknife resampling) are smaller than the size of data points. On the $10\times 10$ lattice, $\Phi/\Phi_0 = 1/8$ and $\Phi/\Phi_0 = 3/8$ can't be achieved exactly, so these values are represented by data obtained with field strengths $\Phi/\Phi_0 = 13/100$ and $\Phi/\Phi_0 = 38/100$, respectively.}
    \label{half-fill-finite-size}
\end{figure}

In \cref{half-fill-finite-size}, we show the  lattice-size dependence of local moment and specific heat for the half-filled Hubbard-Hofstadter model with $U/t=6$. The conclusions that 1) local moment decreases with field and 2) the low-temperature specific heat peak moves to higher temperature are qualitatively true, independent of the lattice size used in DQMC simulations. In general, we expect finite size artifacts to be most severe at small $U$, when our system is least localized, with most extended wavefunctions. As $U$ increases, we expect finite size effects to be reduced, since $U$ localizes electrons, and the system becomes insensitive to boundary conditions. We also expect finite size effects to be most important at low temperatures, when the system is closest to its ground state. Thus, the finite-size effects shown in \cref{half-fill-finite-size} is the worst case for the parameters reported in the main text at half filling, when the system is an AFMI. Even in this worst possible case, on all lattice sizes we simulate, the local moment and specific heat display field dependence trends as reported in the main text. Thus, we believe our finding that an orbital magnetic field delocalizes electrons and weakens the effect of $U$ at half filling is valid in the thermodynamic limit.